\documentclass[bsl,fleqn]{asl}

\usepackage{amssymb, latexsym, cmll}
\usepackage[mathscr]{eucal}
\usepackage{proof, multicol}
\usepackage[graph,curve,dvips,matrix]{xy}

\DeclareMathOperator{\sgn}{sgn}
\DeclareMathOperator{\lab}{lab}

\newtheorem{Def}{Definition} 

\newtheorem{Theorem}[Def]{Theorem}

\newtheorem{Cor}[Def]{Corollary}

\allowdisplaybreaks[2]

\DeclareMathAlphabet{\eufb}{U}{euf}{b}{n}

\title{A survey of proof nets and matrices for substructural logics}

\author{Sean A. Fulop}
\revauthor{Fulop, Sean A.}

\address{Dept.\ of Linguistics\\
California State University, Fresno\\
5245 N. Backer Ave. PB92\\
Fresno, CA 93740-8001}

\email{sfulop@csufresno.edu}

\begin{document}

\begin{abstract}
This paper is a survey of two kinds of ``compressed'' proof schemes,
the \emph{matrix method} and \emph{proof nets,} as applied to a
variety of logics ranging along the substructural hierarchy
 from classical all the way down to the
nonassociative Lambek system.
A novel treatment of proof nets for the latter
is provided. 
Descriptions of proof nets and matrices are given in a uniform
notation based on sequents, so that the properties of the schemes for the
various logics can be easily compared.
\end{abstract}

\maketitle

\footnotetext{Thanks are owed to Greg Restall for commenting on an early version, and to Kosta Do\v{s}en for some important insights.}

\section{Introduction}

This paper provides a survey of two kinds of ``compressed'' proof schemes, the \emph{matrix method} and \emph{proof nets,} as applied to a variety of logics ranging along the substructural hierarchy \cite{Restall2000} from classical all the way down to the nonassociative Lambek system. 
There appears to be a paucity of survey literature that discusses proof nets for a variety of logics in a uniform notation, and even less which discusses matrix methods in relation to proof nets.
It is the author's hope that this paper can provide in one source a host of information and methodology for proof nets and matrices, which could allow further research extending and using these techniques to be more easily conducted.

Section 2 provides the necessary background, reviewing Gentzen's sequent calculus for classical logic (with which readers are assumed
to have some familiarity), presenting the ``unifying notation'' of signed formulae that was introduced by Smullyan \cite{Smullyan1963} for use in his tableaux, and discussing the philosophy of compressed proofs.
Section 3 presents the matrix method, which works for both classical and linear logic, in some detail.
Multiplicative linear logic (MLL) is introduced here using a two-sided sequent calculus that is similar to the classical case.
The matrix method, although originally presented as a compressed proof scheme for formulae only, is extended to operate on two-sided sequents quite readily. 

Section 4 describes proof nets for a variety of logics, beginning with the canonical case of MLL.
The proof nets are defined in a two-sided framework that can be
directly applied to two-sided sequents, so that proof nets for the
various logics can be readily compared. From here it is possible to go
both up and down the substructural hierarchy; after also considering proof nets for classical logic, two versions of the Lambek calculus are treated.
It is observed how the Lambek systems, with their increasingly rigid structural requirements on the layout of the formulae in a sequent, require more strictly geometrical conditions on correct proof nets.
The nonassociative Lambek calculus is here provided with a two-sided
proof net system and a geometric correctness condition for the first time.

\section{Analytic theorem proving}

\subsection{Sequent and tableau systems}

We begin the
discussion with Gentzen's \cite{Gentzen1934} sequent calculus for
classical logic.
This deductive system permits the proof of special \emph{sequent}
statements of the form $\Gamma \Rightarrow \Delta$. 
In a typical notation, $A, B, \ldots$ stand for proposition
occurrences, while $\Theta, \Gamma, \ldots$ stand for sequences of proposition occurrences.
A sequent in classical logic is often interpreted metalogically as a statement that the
(possibly empty) formula sequence $\Delta$, the \emph{succedent,}
follows from the (possibly empty) formula sequence $\Gamma$, the \emph{antecedent,} in a natural deduction or axiomatic system of classical logic. 
To permit this interpretation, the succedent must be
understood as a disjunction of its formulae, while the antecedent must
be understood as a conjunction of its formulae.
The standard (after Gentzen) presentation of the classical sequent calculus involves sequents, as just described, which may have formulae on either side of the arrow; such a presentation is known as a \emph{two-sided} sequent calculus.
Gentzen's rules of inference for the classical sequent calculus may be presented as follows:

\begin{Def}[Classical sequent calculus \cite{Gentzen1934}] \mbox{}\\
\begin{multicols}{2}
\setlength{\abovedisplayskip}{0pt}
\begin{gather*}
D \Rightarrow D\quad \textup{(Axiom)}\\[10pt]
\genfrac{}{}{}{0}{\Gamma \Rightarrow \Theta, A}{\neg A, \Gamma
\Rightarrow \Theta}\quad \textup{($\neg$ L)}\\[10pt]
 \genfrac{}{}{}{0}{A, \Gamma
  \Rightarrow \Theta}{\Gamma \Rightarrow \Theta, \neg A}\quad \textup{($\neg$ R)}\\[10pt]
\genfrac{}{}{}{0}{\Gamma \Rightarrow \Theta, A \quad B, \Delta \Rightarrow
\Lambda}{A \rightarrow B, \Gamma, \Delta \Rightarrow \Theta, \Lambda}\quad \textup{($\rightarrow$ L)}\\[10pt]
 \genfrac{}{}{}{0}{A, \Gamma \Rightarrow \Theta, B}{\Gamma \Rightarrow \Theta, A \rightarrow B}\quad \textup{($\rightarrow$ R)}\\[10pt]
\genfrac{}{}{}{0}{A, \Gamma \Rightarrow \Theta}{A \wedge B, \Gamma \Rightarrow \Theta}\\
 \genfrac{}{}{}{0}{B, \Gamma \Rightarrow \Theta}{A \wedge B, \Gamma \Rightarrow \Theta}\quad \textup{($\wedge$ L)}\\[10pt] 
\genfrac{}{}{}{0}{\Gamma \Rightarrow \Theta, A \quad \Gamma \Rightarrow \Theta, B}{\Gamma \Rightarrow \Theta, A \wedge B}\quad \textup{($\wedge$ R)}\\[10pt] 
\genfrac{}{}{}{0}{A, \Gamma \Rightarrow \Theta \quad B, \Gamma \Rightarrow \Theta}{A \vee B, \Gamma \Rightarrow \Theta}\quad \textup{($\vee$ L)}\\[10pt]
\genfrac{}{}{}{0}{\Gamma \Rightarrow \Theta, A}{\Gamma \Rightarrow \Theta, A \vee B}\\
 \genfrac{}{}{}{0}{\Gamma \Rightarrow \Theta, B}{\Gamma \Rightarrow \Theta, A \vee B}\quad \textup{($\vee$ R)} 
\end{gather*} \end{multicols}
\end{Def}
The above gives the so-called \emph{logical rules} which show how the
operators work.  Because the left and right sides of a sequent are
considered as sequences, to obtain classical logic one also requires
Gentzen's \emph{structural rules}---which are no less logical, in spite of the terms---of Thinning/Weakening (any new proposition may
be added to the antecedent or succedent of a provable sequent),
Contraction (repeated occurrences of the same formula may be
eliminated from antecedent or succedent) and Exchange (any pair of
adjacent formulae may be swapped).
A sequent calculus proof is then a tree-like figure with initial
sequents (possibly axioms)
at the top and a conclusion at the bottom called the
\emph{endsequent.}
To prove a single formula of classical logic, the initial sequents
must be axioms, and the endsequent must have
this formula as the succedent with an empty antecedent.

The only other rule permitted in a sequent calculus is known as ``Cut,'' which certifies a kind of transitivity for sequents:
\[ \genfrac{}{}{}{0}{\Gamma \Rightarrow \Theta, D \qquad D, \Delta \Rightarrow \Lambda}{\Gamma, \Delta \Rightarrow \Theta, \Lambda} \]
Gentzen's important result was his ``Hauptsatz'' stating that the Cut rule can be eliminated with no loss of logical power for the system.
  The resulting Cut-free sequent system then enjoys the \emph{subformula
  property,} meaning that ``the formulae occurring in any [Cut-free] proof of a
given endsequent are all subformulae of the endsequent''
\cite{Wallen1990} (using an obvious definition of \emph{subformula}).
It is plain from inspecting the Cut rule that this cannot be a property
of proofs which use it. 
Thanks to the subformula property, a classical sequent proof search can be
``goal-directed,'' working upward from the endsequent whose proof we
seek by applying the inference rules in reverse.
Of course, only Cut-free proofs can ever be discovered in this fashion.

A goal-directed deduction system is often called \emph{analytic,}
highlighting the idea that one analyzes the goal sequent to construct
(or fail to construct) its proof.
The opposite of this is then called a \emph{synthetic} system,
in which one works from the premisses to the proven expression. A
natural deduction system (e.g.\ \cite{Gentzen1934}) is one example which is usually thought of as synthetic, since it is
not generally used to construct a natural deduction proof
``upwards'' from the conclusion.
It is worth noting, however, that natural deductions in the logics considered here can normally enjoy the subformula property, and so natural deduction can be regarded as closer to an analytic system than it is sometimes given credit for being.
All of the proof methods discussed in this paper are analytic because
our focus is on ``compressed'' proof schemes, and so
the logical systems to be discussed will be limited to their Cut-free
versions. 

Smullyan \cite{Smullyan1968} developed a classical deductive system
called \emph{analytic tab\-leaux} based upon foundations laid by Beth
\cite{Beth1959}.
It is definitionally equivalent to a Cut-free sequent calculus handling sets of
formulae (thus doing without structural rules), but the inference rules are
explicitly turned upside-down, highlighting the analytic approach by
placing the desired goal expression at the top of the proof (called a
tableau).


\subsection{Unifying notation of signed formulae}
\label{unifying}

It would be redundant to present the inference rules of Smullyan's
tableau calculus for classical logic, since they are essentially the
same as those of Gentzen's sequent calculus.
One important element of Smullyan's treatment that will be important
for our discussion throughout, however, is his ``unifying notation''
which uses \emph{signed formulae} classified into two basic varieties \cite{Smullyan1963}. 
A signed formula is just a formula $P$ which is annotated by a sign,
or \emph{polarity,}
which we will show as either $P^+$ or $P^-$. 
The sign is used to indicate the ``negation environment'' of the
formula occurrence within a sequent, so that negative formula occurrences are all those that
are within the scope of an explicit or implicit negation.
An ``implicit negation'' environment is always (and only) the antecedent of
a conditional or of a sequent. 
The above definition comes from the truth-functional equivalence
between formulae $A \rightarrow B$ and $\neg A \vee B$.
 Smullyan used signed formulae to avoid writing sequents
 directly with the sequent arrow; his rules for positive formulae exactly mirror
the succedent (R) rules in the sequent calculus, while tableau rules for negative formulae mirror antecedent (L) rules in the sequent calculus.

Signed formulae are then classified by Smullyan into two fundamental
kinds, which can be determined by inspecting the sequent rules shown
above.
The key question is whether the inference rule ``branches,'' having
two premisses.
A branching rule governs a signed formula which is termed
\emph{disjunctive,} after the canonical example of the rule ($\vee$
L).
A rule with only one premiss, on the other hand, governs a signed
formula which is termed \emph{conjunctive.}
The conjunctive, or \emph{$\alpha$-type} signed formulae in classical logic
are these: 
\[ (A\wedge B)^-, (A \vee B)^+, (A\rightarrow B)^+, (\neg A)^+, (\neg A)^- \]
The disjunctive, or \emph{$\beta$-type} formulae are the others:
\[ (A \vee B)^-, (A \wedge B)^+, (A \rightarrow B)^- \]

\subsection{Proof compression}
A key application of analytic deductive methods has been automated
deduction.
A significant problem for the sequent/tableau systems in this arena is
the inefficiency resulting from a large search space.
Much progress has been made in the development of efficient proof
search by application of goal-directed logic programming methods such
as resolution (e.g.\ \cite{GabbayOlivetti2000}).
The primary focus here will not be on efficient search for complete
proofs, but rather on the question of how can an analytic
proof be compressed, and thereby become a fundamentally
different sort of object that can be viewed in a new way, and possibly constructed more efficiently.

It has been explained many times in the literature (e.g.\ \cite{DyckhoffPinto1999}) that, even restricting attention to
Cut-free proofs, the sequent and tableau calculi may validate numerous
proofs of a given sequent.
These several proofs may differ either ``trivially'' or non-trivially in the order of application of
some of the rules.
In this paper, we will consider two kinds of ``compressed proof objects,'' which differ philosophically with respect to the spurious ambiguities just mentioned. 
The first of these, the \emph{matrix method,} constructs a
minimal compressed proof object that is really nothing beyond a
provability test.
There is a compelling argument that a matrix proof of a sequent is not
really a proof anymore, because not only trivially different, but also nontrivially different sequent/tableau proofs are all collapsed.
A provable sequent has, by definition, exactly one matrix
demonstrating its validity.

The second kind of compressed proof object to be surveyed here is the
\emph{proof net.}
Proof nets were originally described for linear logic
\cite{Girard1987}, and it has been claimed for that system that they compress proofs to
``just the right extent,'' in the sense that any two sequent/tableau
proofs which are \emph{nontrivially different} will have distinct
corresponding proof nets, while any two sequent/tableau proofs which
differ only \emph{spuriously} (i.e.\ by ``harmless'' permutations of the
rules) will have the same proof net corresponding \cite{Strassburger2006}.
The philosophy behind proof nets is to always seek, if not achieve, this correspondence for the logic at hand, because a proof net is supposed to be something beyond a minimal provability test---proponents think of it as ``the essence of a proof.''
There is, however, no current consensus among logicians as to what precisely should count as a \emph{nontrivial} difference between two proofs, and so in the sequel we will need to discuss this matter further in relation to the key logical problem of the \emph{identity of proofs.}

\section{Matrix methods}

\subsection{Classical logic}
In classical and intuitionistic logic, the redundancies and other difficulties with standard proof calculi led to the matrix methods, independently invented by Bibel \cite{Bibel1980} and Andrews \cite{Andrews1981}, but perfected by Wallen \cite{Wallen1990}.
Here we follow Wallen's exposition.

Step one of the matrix method, and also ultimately of the proof net method, is to decompose the target sequent or formula into a tree of its subformulae that keeps track of the signs.
\begin{Def}[\cite{Wallen1990}]
A \emph{formula tree} for a signed formula $A^g, g \in \{+, -\}$ is a
tree of subformulae of $A$ together with an assignment of a sign to each
position $k$ of the formula tree.
Each position $k$ then contains a signed formula \emph{occurrence}
from within $A$; the formula
occurrence apart
from its sign at $k$
is called the \emph{label} of $k$.
Let $\lab(k)$ and $\sgn(k)$ denote the label and sign of position $k$
respectively.
For each position $k$, if $\lab(k)$ occurs positively in $A^g$, then
$\sgn(k) = g$.
If $\lab(k)$ occurs negatively in $A^g$, then $\sgn(k)$ is the
opposite sign from $g$. 
\end{Def}

\begin{Def}[\cite{Wallen1990}]
A \emph{path} through formula $A^g$ is a subset of the positions of its
formula tree, defined inductively:
\begin{enumerate}
\item $\{ k_0 \}$, the root position, is a path.
\item If $s, \alpha$ is a path, so is $(s - \{\alpha\}), \alpha_1,
  \alpha_2$, for conjunctive $\alpha$ having $\alpha_1, \alpha_2$ as
  immediate subformulae.
\item If $s, \alpha$ is a path, so is $(s - \{\alpha\}), \alpha_1$,
  for conjunctive $\alpha$ having $\alpha_1$ as its sole immediate
  subformulae (this is the case where $\alpha$ is a negation).
\item If $s, \beta$ is a path, so is $(s - \{\beta\}), \beta_1$, and
  so is $(s - \{\beta\}), \beta_2$, for disjunctive $\beta$ having
  $\beta_1, \beta_2$ as immediate subformulae.
\end{enumerate}
The second through fourth clauses above can be regarded as path
\emph{reduction} steps.
A completely reduced path will consist entirely of (signed) positions labeled
by atoms, and is called an \emph{atomic} path.
\end{Def}


The above formulation can be easily extended from signed formulae to
two-sided sequents of signed formulae.
The simplest way is to decompose each of the antecedent formulae and
succedent formulae separately.
The antecedent formulae are negatively signed, while the
succedent formulae are positively signed, and one decomposes the
whole set of signed formulae into a set of formula trees as above,
treating the compound tree as a single graph with ``multiple roots.''
The above definitions of a path through the tree and an atomic path can
then be applied mutatis mutandis. 

\begin{Def}[Matrix representation \cite{Wallen1990}]
\mbox{}\\
\begin{enumerate}
\item If signed formula $A^g$ is conjunctive, its matrix representation is a
row matrix having as element(s) the signed component(s) found immediately below in its formula tree.
\item If signed formula $A^g$ is disjunctive, its matrix
  representation is a column matrix having as elements the signed components
  found immediately below in its formula tree.
\item If signed formula $A^g$ is atomic, it is its own matrix
  representation.
\end{enumerate}
\end{Def}
A completed matrix for a signed formula must have every subformula in every submatrix converted to matrix representation; matrices are to be nested as needed.
The matrix representation extends to signed two-sided classical sequents by
a simple adaptation of the procedure described above for sequent
trees.
The matrix of a sequent is then simply a ``row matrix'' whose elements are the respective matrices of the constituent formulae.
This fact can be related to the ``metalogical'' view of a sequent in which the antecedent formulae are conjoined while the succedent formulae are disjoined; observe that a conjunction in the antecedent and a disjunction in the succedent are each formulae of conjunctive type, and so a row matrix is the correct form for each.

Every atomic path through a signed formula (or sequent) is now
represented by the sequence of signed atoms encountered on a left-right sequence of steps through the columns in its completed
matrix---submatrices are to be stepped through as well in this
procedure \cite{Wallen1990}.
When a step enters a column matrix, only one row is selected
(nondeterministically) for the
atomic path, while the other is ignored.

The key idea at the core of the matrix method is that of a spanning set of connections.
\begin{Def}
A \emph{connection} is a pair of atomic positions in some path through
$A^g$, which have the same label but opposite signs.
A given set of connections is said to \emph{span} $A^g$ iff every atomic
path through $A^g$ contains a connection from the set.
\end{Def}

\begin{Theorem}[\cite{Wallen1990}]
For signed propositional formula $A^+$, the existence of a spanning
set of connections for it ensures its provability in classical logic,
and vice versa.
\end{Theorem}

The above definition and theorem concerning a spanning set of
connections also extends in a simple fashion mutatis mutandis to
signed two-sided sequents.
For clarity, this may be stated as follows:
\begin{Cor}
For signed sequent $\Gamma^- \Rightarrow \Delta^+$, the existence of a spanning set of connections for it ensures its provability in classical sequent calculus, and vice versa.
\end{Cor}
An example sequent provable in classical logic is shown in
(\ref{Classicseq1}); the corresponding matrix of the sequent is shown in (\ref{Classicmatrix1}). 
\begin{gather}
\label{Classicseq1}\neg A,\; B \rightarrow A \Rightarrow \neg B\\ 
\label{Classicmatrix1}\begin{bmatrix} [A^+]& \begin{bmatrix} B^+\\ A^- \end{bmatrix} \end{bmatrix}
\quad [B^-]
\end{gather}
This matrix presents two atomic paths: $(A^+, B^+, B^-)$ and $(A^+,
A^-, B^-)$.
The spanning set of connections is then $\{ \langle A^+, A^- \rangle,
\langle B^+, B^- \rangle\}$, the existence of which shows the sequent
to be provable in classical logic.

Any sequent of classical propositional logic can be tested for
provability using our adaptation of Wallen's matrix method to
two-sided sequents; the earlier method due to
Bibel was less general, requiring a goal formula to be given in
negation normal form.

\subsection{Multiplicative linear logic}

Linear logic was introduced by Girard \cite{Girard1987}, and has since been the object of much study.
For our discussion of compressed proof objects, only selected fragments will be used.
We present a two-sided sequent formulation of the \emph{unit-free} multiplicative fragment.
It is two-sided in the previously used sense that the derivable sequents have possibly nonempty antecedent and succedent.
This logic is commonly named MLL$^-$.
\begin{Def}[Sequent calculus for MLL$^-$ \cite{Moot2002}] \mbox{}\\[-14pt]
\begin{multicols}{2}
\setlength{\abovedisplayskip}{0pt}
\begin{gather*}
D \Rightarrow D\quad \textup{(Axiom)}\\[10pt]
\genfrac{}{}{}{0}{\Gamma \Rightarrow A, \Delta}{\Gamma, A^\bot \Rightarrow \Delta}\quad \textup{($^\bot$ L)}\\[10pt]
 \genfrac{}{}{}{0}{\Gamma, A  \Rightarrow \Delta}{\Gamma \Rightarrow A^\bot , \Delta}\quad \textup{($^\bot$ R)}\\[10pt]
\genfrac{}{}{}{0}{\Gamma, A, B \Rightarrow \Delta}{\Gamma, A \otimes B \Rightarrow \Delta}\quad \textup{($\otimes$ L)}\\[10pt]
 \genfrac{}{}{}{0}{\Gamma \Rightarrow A, \Delta \quad \Gamma' \Rightarrow B, \Delta'}{\Gamma, \Gamma' \Rightarrow A \otimes B, \Delta, \Delta'}\quad \textup{($\otimes$ R)}\\[10pt]
\genfrac{}{}{}{0}{\Gamma, A \Rightarrow \Delta \quad \Gamma', B \Rightarrow \Delta'}{\Gamma, \Gamma', A \parr B \Rightarrow \Delta, \Delta'}\quad \textup{($\parr$ L)}\\[10pt]
 \genfrac{}{}{}{0}{\Gamma \Rightarrow A, B, \Delta}{\Gamma \Rightarrow A \parr B, \Delta}\quad \textup{($\parr$ R)}\\[10pt]
\genfrac{}{}{}{0}{\Gamma \Rightarrow A, \Delta \quad \Gamma', B \Rightarrow \Delta'}{\Gamma, \Gamma', A \multimap B \Rightarrow \Delta, \Delta'}\quad \textup{($\multimap$ L)}\\[10pt] 
 \genfrac{}{}{}{0}{\Gamma, A \Rightarrow B, \Delta}{\Gamma \Rightarrow A \multimap B, \Delta}\quad \textup{($\multimap$ R)}
\end{gather*} \end{multicols}
\end{Def}


Linear logic is \emph{substructural,} which means that some of Gentzen's structural rules for classical logic are not allowed.
The only one of Gentzen's structural rules that is allowed now is Exchange, so the sequents in essence keep track of formula occurrences, meaning each side of the sequent constitutes effectively a multiset of occurrences.
We also have the Cut rule allowed, but it is eliminable as before, and
we focus only on the Cut-free version.
There follows an example proof of a sequent in MLL$^-$:
\bigskip

\infer[\textrm{($^\bot$ L)}]{A \parr B, (B \otimes C)^\bot \Rightarrow C \multimap
  A}{\infer[\textrm{($\multimap$ R)}]{A \parr B \Rightarrow C \multimap A, B \otimes
    C}{\infer[\textrm{($\otimes$ R)}]{A \parr B, C \Rightarrow A, B \otimes
      C}{\infer[\textrm{($\parr$ R)}]{A \parr
        B \Rightarrow A, B}{A \Rightarrow A & B \Rightarrow B} & C
      \Rightarrow C}}}
\bigskip

The matrix method for linear logic was presented by Galmiche \cite{Galmiche2000}.
The matrix representation of a sequent is obtained from the signed sequent tree just as with classical logic.
The matrix for the above sequent is shown in (\ref{MLLmatrix}); a spanning set of connections for this matrix is given in (\ref{connections}).

\begin{gather}
\label{MLLmatrix} \begin{bmatrix} A^-\\ B^- \end{bmatrix} \quad \begin{bmatrix} B^+\\
  C^+ \end{bmatrix} \quad \begin{bmatrix} C^- & A^+ \end{bmatrix}\\
\label{connections} \{ \langle A^+, A^-
\rangle, \langle B^+, B^- \rangle, \langle C^+, C^- \rangle \}
\end{gather}

Reflecting the differences between MLL$^-$ and classical logic, it is no longer sufficient that the sequent matrix possesses a spanning set of connections, however.
Galmiche stated the additional requirements for the set $\mathcal{S}$ of connections to \emph{linearly span} a matrix in the following way:
1) All atomic occurrences in the matrix occur exactly once with each polarity in $\mathcal{S}$; 2) no pair of connections in $\mathcal{S}$ has overlapping elements; 3) $\mathcal{S}$ is a minimal spanning set.   
It is easy to see that the set of connections in (\ref{connections}) does linearly span the above
matrix for the sequent.
Galmiche stated the theorem that a sequent of MLL$^-$ is provable just in case its matrix possesses a set of connections which linearly spans it.
It is notable that, despite presenting this as a proven fact, Galmiche never really proved it in his paper.
It is nevertheless possible for us to appreciate, at a glance at least, how the additional conditions defining a linearly spanning set derive from the characteristic that MLL handles multisets of formulae (cf.\ condition 1), effectively keeping track of formula occurrences and not allowing contractions of repeated formulae (cf.\ condition 2) or extraneous formulae (cf.\ condition 3) into a proof.

Having discussed the matrix method and signed formulae, it will be
much easier to understand proof nets, a subject to which we turn next.

\section{Proof net methods}

The sequent and tableau methods yield too many possible proofs of a given sequent, and have an undesirable amount of notational redundancy for automated theorem proving applications.
The matrix method described above has certainly eliminated the redundancy, but now there are in a sense too few proofs of a given sequent for some applications; in fact, each provable sequent has precisely one matrix.
This may be acceptable for theorem proving, but there are theoretical
reasons to desire a proof representation that captures ``the essence
of a proof.''
This notion is related to the general problem of the identity of
proofs \cite{Dosen2003}, and the compressed proof objects known as
\emph{proof nets} have been put forth as solving this problem for MLL,
at least \cite{Strassburger2006}.

\subsection{MLL$^-$}


Our discussion of proof nets must begin with MLL$^-$, since Girard
\cite{Girard1987} developed linear logic and proof nets at the same
time.
The proof nets are in fact a subclass of the decomposition graphs known as proof structures.
Just as the matrix of a sequent is produced from the decomposition of
the signed formulae while distinguishing conjunctive from disjunctive
formulae, a \emph{proof structure} is a special graph that is drawn
from a formula or sequent decomposition, also keeping track of the polarities and
the conjunctive or disjunctive nature of the signed formulae.
The subgraphs which are drawn for each type of formula decomposition
are traditionally called \emph{links}; to complete the proof
structure, pairs of atoms having opposite polarity are linked
together by edges called \emph{axiom links.} 

The MLL$^-$ link schemata which may be
used to decompose formulae in a proof structure are shown below.
It should be mentioned that these link drawings are to be viewed as
representing graphs without further geometric properties, so that the
specific orientation of a link drawing or whether edges cross is unimportant.
A complete proof structure for the provable MLL sequent already
studied above follows the link presentation below.
The resulting graph has two sorts of edges, which serve to distinguish
the conjunctive from the disjunctive binary formula occurrences
(negation is excluded from the conjunctive/disjunctive dichotomy for
this purpose).
The binary links shown with solid lines are the disjunctive formulae,
traditionally called \emph{times} links (typified by the link for the times connective $\otimes$ in a positive context), while the dotted lines show
the conjunctive formulae, traditionally called \emph{par} links (and typified by the link for the par $\parr$ connective in a positive context).
Axiom links are shown with curved lines in the proof structures.
Some formal definitions follow.

\textbf{MLL$^-$ links:}
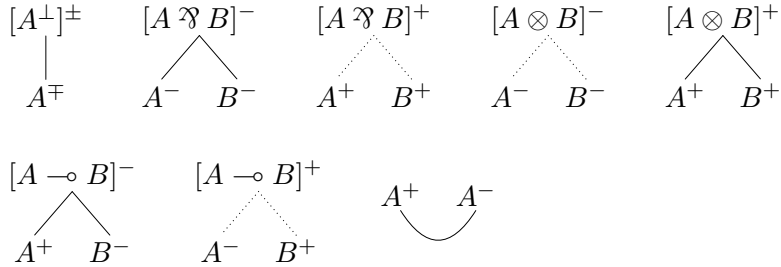
\begin{gather*}\object{\xybox{0*{[A^\bot]^\pm}, +/d1cm/*{A^\mp},
+/u6pt/;0+/d6pt/**@{-} }}
\qquad \object{\xybox{0*{[A \parr B]^-}, +/d1cm/+/l5mm/="A-"*{A^-},
+/r1cm/="B-"*{B^-}
,"A-"+/u6pt/;0+/d6pt/**@{-};"B-"+/u6pt/**@{-}}}
\qquad \object{\xybox{0*{[A \parr B]^+}, +/d1cm/+/l5mm/="A+"*{A^+},
+/r1cm/="B+"*{B^+}
,"A+"+/u6pt/;0+/d6pt/**@{.};"B+"+/u6pt/**@{.}}}
\qquad \object{\xybox{0*{[A \otimes B]^-}, +/d1cm/+/l5mm/="A-"*{A^-},
+/r1cm/="B-"*{B^-}
,"A-"+/u6pt/;0+/d6pt/**@{.};"B-"+/u6pt/**@{.}}}
\qquad \object{\xybox{0*{[A \otimes B]^+}, +/d1cm/+/l5mm/="A+"*{A^+},
+/r1cm/="B+"*{B^+}
,"A+"+/u6pt/;0+/d6pt/**@{-};"B+"+/u6pt/**@{-}}}\\[16pt]
\object{\xybox{0*{[A \multimap B]^-}, +/d1cm/+/l5mm/="A+"*{A^+},
+/r1cm/="B-"*{B^-}
,"A+"+/u6pt/;0+/d6pt/**@{-};"B-"+/u6pt/**@{-}}}
\qquad \object{\xybox{0*{[A \multimap B]^+}, +/d1cm/+/l5mm/="A-"*{A^-},
+/r1cm/="B+"*{B^+}
,"A-"+/u6pt/;0+/d6pt/**@{.};"B+"+/u6pt/**@{.}}}
\qquad \object{\xybox{0*{A^+}, +/r1cm/="A-"*{A^-}, 0+/d6pt/;"A-"+/d6pt/**\crv{0+/r5mm/+/d1cm/} }}
\end{gather*}
Formula(e) on top of each link are called \emph{conclusion,} and formulae on the bottom of a link are called \emph{premiss.}
The axiom link is unique in having no premisses and two conclusions.

\textbf{MLL$^-$ proof structure:}
\begin{Def}[\cite{Moot2002}]
A \emph{proof structure} $\langle S, \mathcal{L}\rangle$ consists of a set $S$ of signed formulae and a set $\mathcal{L}$ of links over $S$ (from the above possibilities).
A proof structure must also satisfy the conditions:
\begin{itemize}
\item Every formula in $S$ is at most once the premiss of a link;
\item Every formula in $S$ is exactly once the conclusion of a link.
\end{itemize}\label{MLLproofstruc}
\end{Def}

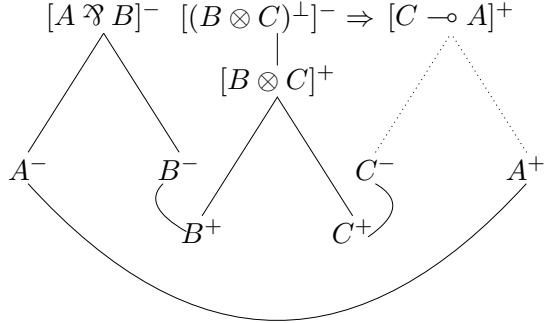
\[\begin{xy}
0*{[A \parr B]^-}, +/r6em/="BtimesCneg"*{[(B \otimes C)^\bot]^- \Rightarrow}, +/r6em/="CimplyA"*{[C \multimap A]^+} 
,"BtimesCneg"+/d24pt/="BtimesC"*{[B \otimes C]^+}
,0+/l1cm/+/d2cm/="A-"*{A^-}, 0+/r1cm/+/d2cm/="B-"*{B^-}
,"BtimesC"+/l1cm/+/d2cm/="B+"*{B^+}, "BtimesC"+/r1cm/+/d2cm/="C+"*{C^+}
,"CimplyA"+/l1cm/+/d2cm/="C-"*{C^-}, "CimplyA"+/r1cm/+/d2cm/="A+"*{A^+}
, 0+/d6pt/;"A-"+/u6pt/**@{-}, 0+/d6pt/;"B-"+/u6pt/**@{-}
, "BtimesC"+/d6pt/;"B+"+/u6pt/**@{-}, "BtimesC"+/d6pt/;"C+"+/u6pt/**@{-}
, "CimplyA"+/d6pt/;"C-"+/u6pt/**@{.}, "CimplyA"+/d6pt/;"A+"+/u6pt/**@{.}
, "BtimesCneg"+/d6pt/;"BtimesC"+/u6pt/**@{-}
, "B-"+/l6pt/+/d6pt/;"B+"+/l6pt/**\crv{"B-"+/l5mm/+/d5mm/}
, "C-"+/d6pt/;"C+"+/r6pt/+/d2pt/**\crv{"C-"+/r5mm/+/d5mm/}
, "A-"+/d6pt/;"A+"+/d6pt/**\crv{"BtimesC"+/d5cm/}
\end{xy}\]

The proof structure exemplified above is \emph{two-sided,}
because it can be created for a sequent with formulae on both sides
of the arrow $\Rightarrow$.
It is possible to enumerate all possible proof structures for any sequent now by decomposing all connectives until we reach the atomic formulae, and then connecting positive to negative atoms using axiom links \cite{Moot2002}.
The two-sided means of presenting a proof structure is, however, not common in literature on linear logic, and has never been fully described in published
literature.\footnote{The presentation here is derived from class
  lecture notes \cite{Strassburger2006}.}
In the literature, MLL proof structures are almost invariably one-sided---meaning they
can be constructed only for a one-sided sequent calculus with empty antecedents.
Moreover, such structures cannot involve the implication or negation operators
overtly as above, because they furthermore do not keep track of the polarities of formulae.
For our purposes, the usual one-sided proof structures for MLL obscure the fundamental
relationship with matrix methods and the unifying notation of signed
formulae, so here we stick with the two-sided dialect.

Completing a proof structure for an MLL sequent is an important step
toward demonstrating provability of the sequent, but it is not
yet sufficient.
A proof structure for a provable sequent is known as a \emph{proof
  net,} and only those structures which satisfy an additional
correctness condition are indeed proof nets.
An impressive list of alternative correctness conditions for MLL proof
nets has arisen from years of research on the topic, beginning
with the original ``long trip'' condition of Girard \cite{Girard1987}.
This condition is somewhat cumbersome for our purpose here, so we will
first describe a popular correctness condition due to Danos and
Regnier \cite{DanosRegnier1989}.

Take a proof structure as a graph; let us call it $\mathcal{P}$.
Now, let $\sigma(\mathcal{P})$ be a new graph derived from
$\mathcal{P}$ by deleting some edges.
Specifically, $\sigma$ acts to
delete one edge nondeterministically from each par link in
$\mathcal{P}$, and is called a \emph{DR switching.}
\begin{Theorem}[Danos-Regnier correctness condition]
A proof structure $\mathcal{P}$ is a proof net if and only if every DR
switching $\sigma(\mathcal{P})$ of it yields a connected acyclic
graph.
\end{Theorem}

The Danos-Regnier switching condition is easy to apply to small proof
structures---the structure presented above is easily seen to satisfy it---but has
exponential complexity because of the need to check the result of
every DR switching of a proof structure for acyclicity and
connectedness \cite{Moot2002}.

A more efficient condition was first presented in the PhD dissertation
of Danos \cite{Danos1990}, and involves transforming a candidate proof
structure by graph contractions (v.\ \cite{GrossTucker1987} for a formal
definition of graph contraction).
The two Danos contraction rules are presented as follows in \cite{Moot2002}:
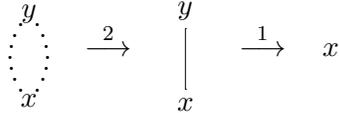
\[
\object{\xybox{0*{y};+/d3em/*{x}**\crv{~**{.}0+/l5mm/+/d1.5em/}**\crv{~**{.}0+/r5mm/+/d1.5em/}} } \quad \overset{2}{\longrightarrow} \quad \object{\xybox{\xymatrix{y \ar@{-}[d] \\ x}} } \quad \overset{1}{\longrightarrow} \quad x 
\]
The basic idea is that two `par' edges transform to a single `times'
edge just if they connect the same two vertices (this can only result
from previous contractions), and any two vertices
linked by a `times' edge contract to one vertex.
\begin{Theorem}[Danos contraction condition]
A proof structure is a proof net if and only if it contracts to a
single vertex by successive application of the Danos contraction rules
above.
\end{Theorem}
The figure sequence below shows the successive contraction of the proof
structure presented above; formulae are irrelevant for this condition,
and are replaced by simple vertex labels.
The equivalence between the Danos-Regnier switching condition and the
Danos contraction condition was proven very simply in
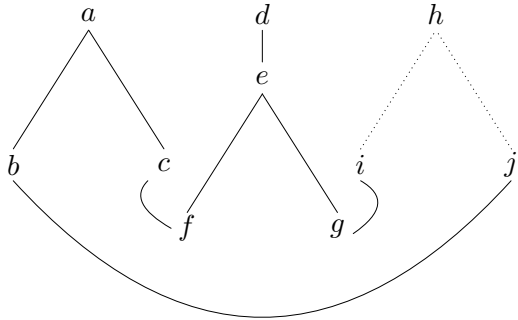
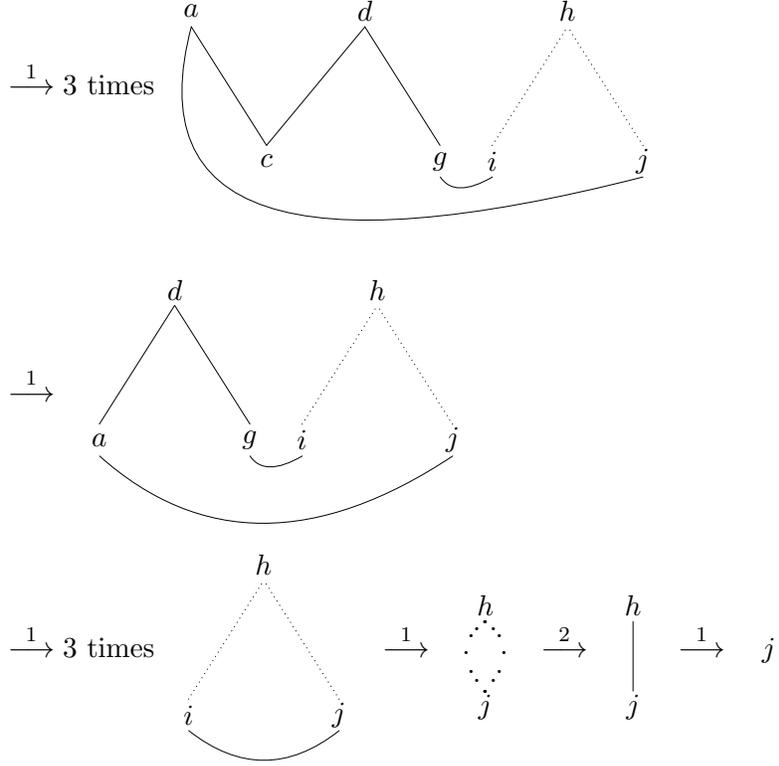
\cite{Strassburger2006}.

\[\begin{xy}
0*{a}, +/r6em/="d"*{d}, +/r6em/="h"*{h} 
,"d"+/d24pt/="e"*{e}
,0+/l1cm/+/d2cm/="b"*{b}, 0+/r1cm/+/d2cm/="c"*{c}
,"e"+/l1cm/+/d2cm/="f"*{f}, "e"+/r1cm/+/d2cm/="g"*{g}
,"h"+/l1cm/+/d2cm/="i"*{i}, "h"+/r1cm/+/d2cm/="j"*{j}
, 0+/d6pt/;"b"+/u6pt/**@{-}, 0+/d6pt/;"c"+/u6pt/**@{-}
, "e"+/d6pt/;"f"+/u6pt/**@{-}, "e"+/d6pt/;"g"+/u6pt/**@{-}
, "h"+/d6pt/;"i"+/u6pt/**@{.}, "h"+/d6pt/;"j"+/u6pt/**@{.}
, "d"+/d6pt/;"e"+/u6pt/**@{-}
, "c"+/l6pt/+/d6pt/;"f"+/l6pt/**\crv{"c"+/l5mm/+/d5mm/}
, "i"+/d6pt/;"g"+/r6pt/+/d2pt/**\crv{"i"+/r5mm/+/d5mm/}
, "b"+/d6pt/;"j"+/d6pt/**\crv{"e"+/d5cm/}
\end{xy}\]

\[ \overset{1}{\longrightarrow} \text{3 times} \quad \object{\xybox{
0*{a}, +/r6em/="d"*{d}, +/r7em/="h"*{h}
,0+/r1cm/+/d2cm/="c"*{c}, "d"+/r1cm/+/d2cm/="g"*{g}
,"h"+/l1cm/+/d2cm/="i"*{i}, "h"+/r1cm/+/d2cm/="j"*{j}
, 0+/d6pt/;"c"+/u6pt/**@{-}
, "d"+/d6pt/;"c"+/u6pt/**@{-}, "d"+/d6pt/;"g"+/u6pt/**@{-}
, "h"+/d6pt/;"i"+/u6pt/**@{.}, "h"+/d6pt/;"j"+/u6pt/**@{.}
, "i"+/d6pt/;"g"+/d6pt/**\crv{"i"+/l5mm/+/d5mm/}
, 0+/d6pt/;"j"+/d6pt/**\crv{"d"+/d4cm/,"c"+/d2cm/+/l2cm/} }}
\]
\bigskip

\[ \overset{1}{\longrightarrow} \quad \object{\xybox{
0="d"*{d}, +/r7em/="h"*{h}
, "d"+/l1cm/+/d2cm/="a"*{a}, "d"+/r1cm/+/d2cm/="g"*{g}
, "h"+/l1cm/+/d2cm/="i"*{i}, "h"+/r1cm/+/d2cm/="j"*{j}
, "d"+/d6pt/;"a"+/u6pt/**@{-}, "d"+/d6pt/;"g"+/u6pt/**@{-}
, "h"+/d6pt/;"i"+/u6pt/**@{.}, "h"+/d6pt/;"j"+/u6pt/**@{.}
, "i"+/d6pt/;"g"+/d6pt/**\crv{"i"+/l5mm/+/d5mm/}
, "a"+/d6pt/;"j"+/d6pt/**\crv{"g"+/d2cm/,} }}
\]
 
\[ \overset{1}{\longrightarrow} \text{3 times} \quad \object{\xybox{
0="h"*{h}, +/l1cm/+/d2cm/="i"*{i}, "h"+/r1cm/+/d2cm/="j"*{j}
, "h"+/d6pt/;"i"+/u6pt/**@{.}, "h"+/d6pt/;"j"+/u6pt/**@{.}
, "i"+/d6pt/;"j"+/d6pt/**\crv{"h"+/d3cm/} }}
\quad \overset{1}{\longrightarrow} \quad \object{\xybox{0*{h};+/d3.5em/="j"*{j},
0+/d6pt/;"j"+/u6pt/**\crv{~**{.}0+/l5mm/+/d1.5em/}**\crv{~**{.}0+/r5mm/+/d1.5em/}} }
\quad \overset{2}{\longrightarrow} \quad \object{\xybox{0*{h}, +/d3.5em/="j"*{j},
0+/d6pt/;"j"+/u6pt/**@{-} }} \quad \overset{1}{\longrightarrow} \quad j
\]

\subsection{Complexity issues}

While complexity of proof methods is not our focus here it must at least be mentioned, since the improvement in efficiency offered by compressed proof objects is a major reason for their promotion and study.
The complexity of the decision problem in MLL has been shown to be NP-complete \cite{Kanovich1991}, so no theorem-proving scheme can ever really be tractable.
The best that can be hoped for is minimal intractability.
The Danos contraction condition as described above has complexity $O(n^2)$ in the size of the
proof structure \cite{Moot2002}.
Guerrini \cite{Guerrini1999}, however, showed how to
convert the contraction algorithm into one with linear complexity.
Another way of developing a linear time correctness check was shown by
Murawski and Ong \cite{MurawskiOng2000}.
  Given the existence of linear-time algorithms to check correctness of a proof structure, the origin of the overall NP complexity must be the sheer number of possible proof structures to be checked, because how to create axiom links can be indeterminate after expanding the formula tree.

Turning to the matrix method, the way to check correctness of a matrix involves traversing all paths through it to see whether there exists a (linearly) spanning set of connections for it.
Now, while a matrix appears to be a proof object of a truly minimal size and graphic intricacy, the worst-case complexity of checking a matrix would seem to be exponential, on the order of $2^n$ in the length of the formula.
This can happen in the case of a formula that involves nested disjunctive subformulae, which will yield nested column matrices through which all paths must be traversed.
For this reason, the matrix method was dismissed out of hand by Hughes \cite{Hughes2006} as not even a ``proof system,'' which has occasionally been defined \cite{CookReckhow1979} as a system in which proofs can at least be checked, if not constructed, in polynomial time. 
The matrix method does bear the singular feature that actually constructing the proof object is deterministic and linear-time.
But, to borrow a common adage, if logic were that easy everyone would be doing it.
The matrix method's powerfully simple proof construction leaves a large debt to be paid on the other end of the deal, when the time comes to check it.
In practical applications, of course, all these considerations are less important than the ultimate competition among the average-case complexities, and discussing that is beyond our scope here.

\subsection{Identity of proofs}

The \emph{identity of proofs} problem remains a significant open research question in logic and proof theory \cite{Dosen2003}.\footnote{This subsection owes a great debt to personal communications with Kosta Do\v{s}en, and I herein communicate some of his arguments.}
Simply put, for any given logical system this is the question of when two apparently different proofs (of the same formula) ought to be regarded as fundamentally ``identical.''
While at first glance this issue appears tangential to the main track of the present paper, it has to be addressed because it has so often been a central concern in the community researching proof nets for the system MLL, among others.
Proof nets have usually been promoted as addressing this question directly \cite{Girard1987}, and have sometimes been claimed to actually solve the issue \cite{Strassburger2006}.
It will now be explained how such claims should be viewed as exaggerated.

There may in general be more than one proof net for a
provable sequent, however there can often be fewer possible proof nets than possible sequent proofs, even in a Cut-free system.
Each proof net has often been viewed as representing an equivalence class of
sequent proofs modulo ``spurious ambiguity,'' while distinct proof nets
will sequentialize respectively to full sequent proofs which are ``nontrivially''
distinct \cite{Strassburger2006}.
In lecture notes (op.\ cit.), Stra{\ss}burger goes so far as to claim a theorem stating that two sequent calculus proofs in MLL translate to the same proof net iff they can be transformed into each other using only ``trivial rule permutations.''
Yet, such a theorem seems to be circular, for in order to have this result one must assert in advance precisely what kinds of sequent rule permutations are held to be trivial and which are nontrivial.
But it is this last issue that remains fundamentally a matter of debate!

Moreover (as Do\v{s}en pointed out to me), on Stra{\ss}burger's analysis, two sequent proofs which differ only by the presence of a useless Cut rule must be held to be nontrivially distinct, because the one with Cut will translate to a proof net involving a Cut link.
Yet there is broad agreement among logicians that a sequent proof involving Cut should be regarded as ``identical'' to its Cut-free variant.
Proof nets, therefore, should not be seen to have solved the identity of proofs question for any logical system. 
As for the matrix of a sequent, there can be only one, so as a proof-theoretical object it does not address the fundamental question of ``identity of proofs'' other than to trivialize it.

\subsection{Intuitionistic MLL}

We refer back to the sequent calculus rules for MLL$^-$; this system is rendered intuitionistic by endowing it with the single-conclusion property, by which all sequents must have just one succedent formula \cite{Moot2002}.
The positive fragment of this system with only
$\otimes$ and $\multimap$ is known in the literature as multiplicative \emph{intuitionistic} linear logic (MILL), and it has some thinly disguised early roots.

A kind of decomposition graph for MILL formulae
was published by Kelly and Mac Lane in 1971 \cite{KellyMacLane1971} in
their study of coherence in categories,
and is possibly the first work on a compressed proof object showing aspects of the
matrix and proof net methods.
The Kelly-MacLane graph of a MILL formula shows its decomposition to signed atoms; if linking atoms in opposite polarity pairs can be achieved, then one has essentially a proof structure, but a correctness criterion is still required for such a structure to be a proof net \cite{Moot2002}.

To build a proof structure in MILL, one begins as
in MLL by decomposing the sequent into subformulae down to the atoms
while keeping track of the polarities and the conjunctive/disjunctive property of the formula at each stage.
The antecedent formulae are first given a negative sign while the
succedent formula is given a positive sign.
Beyond this there are just two operators, and the decomposition
proceeds so that $[X \otimes Y]^\pm$ yields $X^\pm, Y^\pm$, while $[X
\multimap Y]^\pm$ yields $X^\mp, Y^\pm$, as above in MLL.
Signed formulae of the form $[X \otimes Y]^-, [X \multimap Y]^+$ are
the conjunctive ones as in MLL, which are assigned a par link.
The formal definition of a proof structure in MILL (without units) is the same as the above for MLL$^-$ mutatis mutandis, and either of the above correctness conditions for MLL proof nets carries
over to the case of MILL \cite{Moot2002}.

Below we show two proof structures for posited sequents of MILL; only
the first one is a correct proof net, in which each DR switching
yields a connected acyclic graph.
The second proof structure has a cycle following removal of the left
branch of the par link, demonstrating the posited sequent to be
underivable in MILL.
We see that MILL proof nets are merely a subspecies of MLL nets, however, one reason to discuss this logic separately here is to highlight the
much earlier literature \cite{KellyMacLane1971} that first defined
proof structures for this system, and was also first to make use of
signed formulae in a linear logic system.
MILL is in a sense also the archetypal logic in this family possessing
the single-conclusion property.

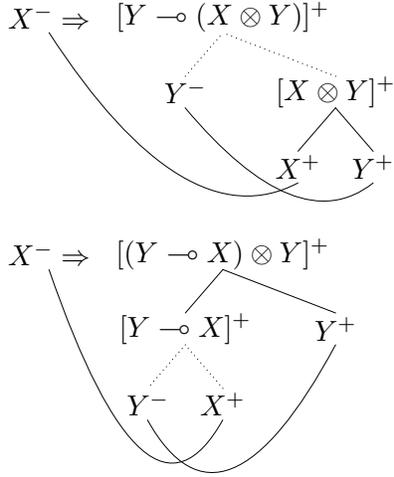
\[\begin{xy}
0*{X^- \Rightarrow}, +/r6em/="succ"*{[Y \multimap (X \otimes Y)]^+}
,+/d1cm/+/l5mm/="Y-"*{Y^-}, +/r2cm/="XtimesY"*{[X \otimes Y]^+}
,+/d1cm/+/l5mm/="X+"*{X^+}, +/r1cm/="Y+"*{Y^+}
,"succ"+/d6pt/;"Y-"+/u6pt/**@{.}, "succ"+/d6pt/;"XtimesY"+/u6pt/**@{.}
,"X+"+/u6pt/;"XtimesY"+/d6pt/**@{-};"Y+"+/u6pt/**@{-}
,0+/d6pt/;"X+"+/d6pt/**\crv{"Y-"+/d2cm/}
,"Y-"+/d6pt/;"Y+"+/d6pt/**\crv{"X+"+/d1cm/}
\end{xy}\]

\[ \begin{xy}
0*{X^- \Rightarrow}, +/r6em/="succ"*{[(Y \multimap X) \otimes Y]^+}
,+/d1cm/+/l5mm/="YimpliesX"*{[Y \multimap X]^+}, +/r2cm/="Y+"*{Y^+}
,"YimpliesX"+/d1cm/+/l5mm/="Y-"*{Y^-}, +/r1cm/="X+"*{X^+}
,"succ"+/d6pt/;"YimpliesX"+/u6pt/**@{-},
"succ"+/d6pt/;"Y+"+/u6pt/**@{-}
,"Y-"+/u6pt/;"YimpliesX"+/d6pt/**@{.};"X+"+/u6pt/**@{.}
,0+/d6pt/;"X+"+/d6pt/**\crv{"Y-"+/d2cm/}
,"Y-"+/d6pt/;"Y+"+/d6pt/**\crv{"X+"+/d2cm/}
\end{xy}\]
\bigskip

\subsection{Classical logic}

Classical (propositional) logic is actually quite similar to linear logic; all of the
differences derive from the presence of Gentzen's structural
rules of Weakening and Contraction.
It is interesting to see how the definition of a proof net carries
over to this case.
Proof nets for classical logic have been developed by Robinson
\cite{Robinson2003} following the two-sided paradigm given above for
linear logic, in which there are distinct links for decomposing each
connective in a positive versus a negative context.
The system therefore derives naturally from Smullyan's unifying
notation for classical logic, although Robinson did not reveal any
awareness of this prior literature.
The conjunctive and disjunctive links for the decomposition of signed
formulae are very similar to the ones needed for MLL, and are presented below with adjustments to suit our notation here (leaving aside the degenerate links which would handle the true and false units, not used here).

\textbf{Classical logic links:}
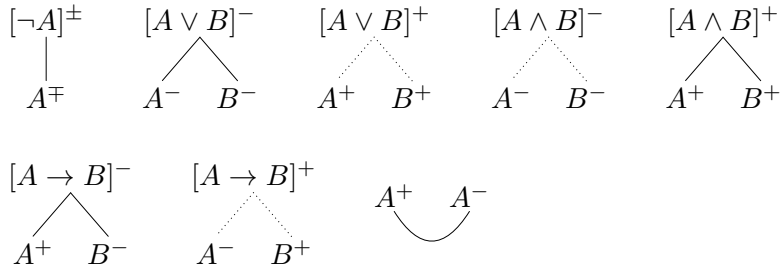
\begin{gather*}\object{\xybox{0*{[\neg A]^\pm}, +/d1cm/*{A^\mp},
+/u6pt/;0+/d6pt/**@{-} }}
\qquad \object{\xybox{0*{[A \vee B]^-}, +/d1cm/+/l5mm/="A-"*{A^-},
+/r1cm/="B-"*{B^-}
,"A-"+/u6pt/;0+/d6pt/**@{-};"B-"+/u6pt/**@{-}}}
\qquad \object{\xybox{0*{[A \vee B]^+}, +/d1cm/+/l5mm/="A+"*{A^+},
+/r1cm/="B+"*{B^+}
,"A+"+/u6pt/;0+/d6pt/**@{.};"B+"+/u6pt/**@{.}}}
\qquad \object{\xybox{0*{[A \wedge B]^-}, +/d1cm/+/l5mm/="A-"*{A^-},
+/r1cm/="B-"*{B^-}
,"A-"+/u6pt/;0+/d6pt/**@{.};"B-"+/u6pt/**@{.}}}
\qquad \object{\xybox{0*{[A \wedge B]^+}, +/d1cm/+/l5mm/="A+"*{A^+},
+/r1cm/="B+"*{B^+}
,"A+"+/u6pt/;0+/d6pt/**@{-};"B+"+/u6pt/**@{-}}}\\[16pt]
\object{\xybox{0*{[A \rightarrow B]^-}, +/d1cm/+/l5mm/="A+"*{A^+},
+/r1cm/="B-"*{B^-}
,"A+"+/u6pt/;0+/d6pt/**@{-};"B-"+/u6pt/**@{-}}}
\qquad \object{\xybox{0*{[A \rightarrow B]^+}, +/d1cm/+/l5mm/="A-"*{A^-},
+/r1cm/="B+"*{B^+}
,"A-"+/u6pt/;0+/d6pt/**@{.};"B+"+/u6pt/**@{.}}}
\qquad \object{\xybox{0*{A^+}, +/r1cm/="A-"*{A^-}, 0+/d6pt/;"A-"+/d6pt/**\crv{0+/r5mm/+/d1cm/} }}
\end{gather*}

As Robinson showed, more is needed to obtain a kind of proof net that
enjoys the same correctness conditions which govern MLL.
Specifically, Robinson added links corresponding to the
structural rules of Contraction and Weakening; the former are conjunctive while the latter are disjunctive.
Once again our presentation changes the link notation somewhat to make it uniform with the presentation of MLL.

\textbf{Structural links:}
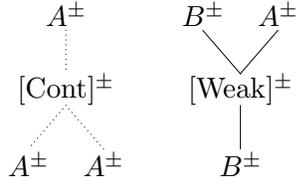
\begin{gather*}\object{\xybox{0*{[\textrm{Cont}]^\pm}, +/u1cm/="u"*{A^\pm}
, 0+/d1cm/+/l5mm/="dl"*{A^\pm}, +/r1cm/="dr"*{A^\pm}, 
"dl"+/u6pt/;0+/d6pt/**@{.};"dr"+/u6pt/**@{.}, 0+/u6pt/;"u"+/d6pt/**@{.} }}
\qquad \object{\xybox{0*{[\textrm{Weak}]^\pm}, +/d1cm/="d"*{B^\pm}
, 0+/u1cm/+/l5mm/="ul"*{B^\pm}, +/r1cm/="ur"*{A^\pm},
"ul"+/d6pt/;0+/u6pt/**@{-};"ur"+/d6pt/**@{-}, 0+/d6pt/;"d"+/u6pt/**@{-} }}
\end{gather*}
With these additions, a proof structure can be constructed for a
classical sequent, relying on Definition \ref{MLLproofstruc} from the MLL case.
The correctness conditions it must meet to be a
proof net for a provable sequent are also carried over from MLL with no
changes.
An example is now shown.

\textbf{Classical proof net:}
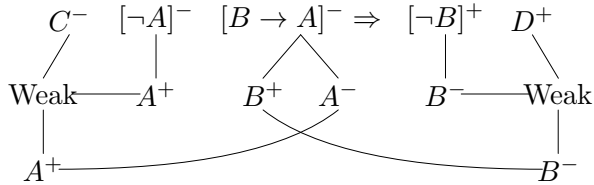
\[
\object{\xybox{0*{C^-}, +/r3em/="Aneg"*{[\neg A]^-},
+/r5em/="BimpliesA"*{[B \rightarrow A]^- \Rightarrow},
+/r5em/="Bneg"*{[\neg B]^+}, +/r3em/="D+"*{D^+}
,"Aneg"+/d1cm/="A+"*{A^+}, "A+"+/l1.5cm/="WeakL"*{\text{Weak}}
,"WeakL"+/d1cm/="A+ax"*{A^+}, "BimpliesA"+/d1cm/+/l5mm/="B+"*{B^+}
,"BimpliesA"+/d1cm/+/r5mm/="A-"*{A^-}, "Bneg"+/d1cm/="B-"*{B^-}
,"B-"+/r1.5cm/="WeakR"*{\text{Weak}}, "WeakR"+/d1cm/="B-ax"*{B^-}
,0+/d6pt/;"WeakL"+/u6pt/**@{-}, "WeakL"+/r1em/;"A+"+/l6pt/**@{-}
,"Aneg"+/d6pt/;"A+"+/u6pt/**@{-}, "B+"+/u6pt/;"BimpliesA"+/d6pt/**@{-};"A-"+/u6pt/**@{-}
,"Bneg"+/d6pt/;"B-"+/u6pt/**@{-}
,"B-"+/r6pt/;"WeakR"+/l1em/**@{-}, "WeakR"+/u6pt/;"D+"+/d6pt/**@{-}
,"WeakL"+/d6pt/;"A+ax"+/u6pt/**@{-}, "WeakR"+/d6pt/;"B-ax"+/u6pt/**@{-}
,"A+ax"+/r6pt/;"A-"+/d6pt/**\crv{"B+"+/d1cm/}
,"B+"+/d6pt/;"B-ax"+/l6pt/**\crv{"A-"+/d1cm/}
}}
\]
This classical proof net turns out to have no conjunctive links, so
it has to be a connected acyclic graph as it is shown according to the Danos-Regnier condition, and indeed it is.
Robinson also gave an elegant, simple explanation connecting this
correctness condition to the unifying notation, to be restated now.
If a proof net comes from a proof, the graph must be simply connected, which forces the switching condition in the following way.
A disjunctive (`times') signed subformula $A \circ B$ for any operator
$\circ$ has ``branched'' premisses which come from separate subproofs,
and so are not yet connected, so the occurrence of $A \circ B$ must be
joined to both premisses otherwise the proof net would end up disconnected.
On the other hand, a conjunctive (`par') signed subformula $C \circ D$ has premisses coming from the same subproof, so they are already connected.
The formula $C \circ D$ must then be joined to exactly one premiss or the graph will contain a cycle.
This explanation is also applicable to the linear logic cases.
It is interesting that the only real difference between the MLL$^-$ and classical proof net systems is the addition of the special links for Contraction and Weakening.

\subsection{Associative Lambek calculus}

So far we have discussed classical logic, which in essence handles formulae in sets, and two varieties of linear logic, which remove the Weakening and Contraction rules, and thereby keep track of occurrences of formulae.
It is useful to note at this juncture that these logics have both
matrix and proof net methods available for checking provability of
sequents, neither of whose correctness conditions refer crucially to the
geometrical arrangement of the proof object. 
It is apparent that a matrix of a sequent does not have any
interesting geometrical properties; moreover, although a proof net is a kind of graph,
there is nothing very ``geometrical'' about these proof nets so far.
It is unimportant whether the link lines in a drawing of the graph cross, for example.
 
Now, we discuss other substructural logics which take away more of the
structural rules, both explicit and implicit.
An important motivation for these logics is found in linguistics,
where they serve as fundamental systems within theories of
``categorial grammar'' and its extension to ``type-logical grammar'' \cite{Morrill1994,Fulop2004a}. 
Our first example is a logic that was first introduced as a ``syntactic calculus'' operating on formulae that were interpreted as linguistic parts of speech \cite{Lambek1958}.
In this guise it is known as the (associative) Lambek calculus.

\begin{Def}[Lambek sequent calculus, \cite{Lambek1958}] \mbox{}\\[-14pt]
\begin{multicols}{2}
\setlength{\abovedisplayskip}{0pt}
\begin{gather*}
D \Rightarrow D\quad \textup{(Axiom)}\\[10pt]
\genfrac{}{}{}{0}{\Delta \Rightarrow B \quad \Gamma[A] \Rightarrow C}{\Gamma[A/B, \Delta] \Rightarrow C}\quad \textup{(/L)}\\[10pt]
\genfrac{}{}{}{0}{\Gamma, B \Rightarrow A}{\Gamma \Rightarrow A/B}\quad \textup{(/R)}\\[10pt]
\genfrac{}{}{}{0}{\Delta \Rightarrow B \quad \Gamma[A] \Rightarrow C}{\Gamma[\Delta, B\backslash A] \Rightarrow C}\quad \textup{($\backslash$L)} \\[10pt]
\genfrac{}{}{}{0}{B, \Gamma \Rightarrow A}{\Gamma \Rightarrow B\backslash A}\quad \textup{($\backslash$R)}\\[10pt]
\genfrac{}{}{}{0}{\Gamma[A, B] \Rightarrow C}{\Gamma[A \bullet B] \Rightarrow C}\quad \textup{($\bullet$L)} \\[10pt]
\genfrac{}{}{}{0}{\Gamma \Rightarrow A \quad \Delta \Rightarrow B}{\Gamma, \Delta \Rightarrow A \bullet B}\quad \textup{($\bullet$R)}\\[10pt]
\genfrac{}{}{}{0}{\Delta \Rightarrow A \quad \Gamma[A] \Rightarrow C}{\Gamma[\Delta] \Rightarrow C}\quad \textup{(Cut)}
\end{gather*} \end{multicols}
\end{Def}

Lambek calculus (notated simply L, or L$_\epsilon$ when empty antecedents are permitted) is a positive logic in which none of
Weakening, Contraction, or Exchange are permitted, so the logical
consequence relation involves sequences of formulae.  
Thus we introduced the
standard notation $\Gamma[\cdot]$, which means a formula sequence with
a place identified for substitution which is matched by another use of
the similar notation in the same inference rule.
Once again, the logic enjoys Cut elimination so we deal solely with
the Cut-free version.
Associativity of the sequences is assumed, as a kind of implicit structural rule (we show what happens in the sequel when even this is removed).

In fact, L$_\epsilon$ is basically MILL without Exchange.
The lack of Exchange (or ``commutativity'') has effectively split the linear implication into a pair of directionally sensitive implications notated with the slash operators.
Each slash is interpreted as saying that the formula on top of the slash results when the formula under it is adjacent on that side.
The newfound sensitivity of the logic to the arrangement of formulae
in a sequence has a profound effect on the definition of a compressed
proof procedure.
Below, the binary links for proof nets in the Lambek system L are provided,
following Roorda \cite{Roorda1991,Roorda1992}; this time, however, the
geometry of the drawings as shown provides important information.
The left-right arrangement of the subformulae in a decomposition link is
now critical; one must swap the order of the subformulae with respect to
the parent formula \emph{in the positive links only.}\footnote{One of
  the very few sources to provide these link drawings for Lambek proof
  nets \cite{Moot2002} has got this condition backwards, unfortunately.}


\textbf{Links for L$_\epsilon$:}
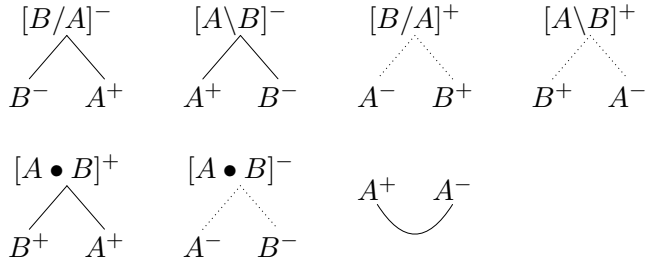
\begin{gather*}
\object{\xybox{0*{[B / A]^-}, +/d1cm/+/l5mm/="B-"*{B^-},
+/r1cm/="A+"*{A^+}
,"A+"+/u6pt/;0+/d6pt/**@{-};"B-"+/u6pt/**@{-}}}
\qquad \object{\xybox{0*{[A \backslash B]^-}, +/d1cm/+/l5mm/="A+"*{A^+},
+/r1cm/="B-"*{B^-}
,"A+"+/u6pt/;0+/d6pt/**@{-};"B-"+/u6pt/**@{-}}}
\qquad \object{\xybox{0*{[B / A]^+}, +/d1cm/+/l5mm/="A-"*{A^-},
+/r1cm/="B+"*{B^+}
,"A-"+/u6pt/;0+/d6pt/**@{.};"B+"+/u6pt/**@{.}}}
\qquad \object{\xybox{0*{[A \backslash B]^+}, +/d1cm/+/l5mm/="B+"*{B^+},+/r1cm/="A-"*{A^-}
,"A-"+/u6pt/;0+/d6pt/**@{.};"B+"+/u6pt/**@{.}}}\\[14pt]
\object{\xybox{0*{[A \bullet B]^+}, +/d1cm/+/l5mm/="B+"*{B^+},
+/r1cm/="A+"*{A^+}
,"A+"+/u6pt/;0+/d6pt/**@{-};"B+"+/u6pt/**@{-}}}
\qquad \object{\xybox{0*{[A \bullet B]^-}, +/d1cm/+/l5mm/="A-"*{A^-},
+/r1cm/="B-"*{B^-}
,"B-"+/u6pt/;0+/d6pt/**@{.};"A-"+/u6pt/**@{.}}}
\qquad \object{\xybox{0*{A^+}, +/r1cm/="A-"*{A^-}, 0+/d6pt/;"A-"+/d6pt/**\crv{0+/r5mm/+/d1cm/} }}
\end{gather*}

\textbf{L proof structure}
It seems that the formal definition of a proof structure in L can be kept the same as for the systems above.  An example is now shown.
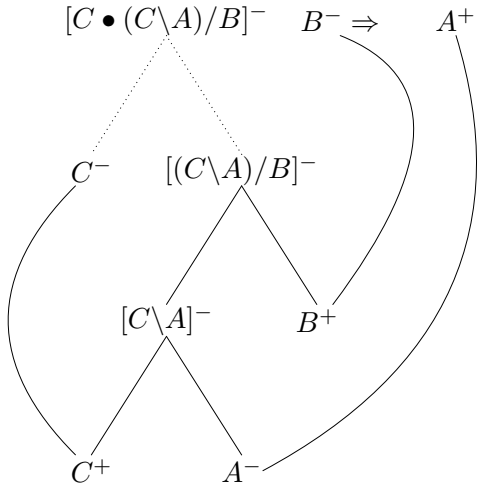
\[\begin{xy}
0*{[C \bullet (C \backslash A)/B]^-}, +/r6em/="B-"*{B^- \Rightarrow}, +/r4em/="A+"*{A^+}
,0+/l1cm/+/d2cm/="C-"*{C^-}, 0+/r1cm/+/d2cm/="CbackA/B"*{[(C\backslash A)/B]^-}
,"CbackA/B"+/l1cm/+/d2cm/="CbackA"*{[C\backslash A]^-}
,"CbackA/B"+/r1cm/+/d2cm/="B+"*{B^+}
,"CbackA"+/l1cm/+/d2cm/="C+"*{C^+}, "CbackA"+/r1cm/+/d2cm/="A-"*{A^-}
,"C-"+/u6pt/;0+/d6pt/**@{.};"CbackA/B"+/u6pt/**@{.}
,"CbackA"+/u6pt/;"CbackA/B"+/d6pt/**@{-};"B+"+/u6pt/**@{-}
,"C+"+/u6pt/;"CbackA"+/d6pt/**@{-};"A-"+/u6pt/**@{-}
,"C-"+/d6pt/+/l6pt/;"C+"+/u6pt/+/l6pt/**\crv{"CbackA"+/l3cm/}
,"B-"+/d6pt/;"B+"+/u6pt/+/r6pt/**\crv{"B-"+/d1cm/+/r2cm/}
,"A+"+/d6pt/;"A-"+/r8pt/**\crv{"B+"+/r3cm/}
\end{xy}\]

This example is
in fact also a proof net for the provable sequent. 
This proof net satisfies the Danos-Regnier condition plus an
additional requirement of planarity which was first proven necessary
by Roorda \cite{Roorda1991}; each DR-switching graph is not only
acyclic connected, but also planar as shown in the
drawing.

Although this treatment here applies generally to the system
L$_\epsilon$ allowing empty antecedents, it has been shown
\cite{LamarcheRetore1996} that we may exclude all sequents with empty
antecedent by an additional requirement about \emph{subnets} of a
proof net.
A subnet is, in our notation, a down-closed subset of the nodes such that axiom links
stay inside the substructure.
Then, to exclude sequents with empty antecedent, it is sufficient to
require that every subnet of a proof net possess at least
 two conclusions (i.e.\ local root formulae at the top).
 
A  different presentation of a noncommutative linear logic was also
  shown to require planar proof nets \cite{Abrusci1991}, around the
  same time as Roorda's result about the Lambek system. 
The NCMLL system described in the reference is equivalent in its
multiplicative fragment to another noncommutative logic \cite{Pentus1997b}, which in turn
is a conservative extension of L$_\epsilon$ \cite{Abrusci1997}.
Thus it is beginning to look as if noncommutativity of the logic (i.e.\ lack of
Exchange) leads directly to a new geometric requirement of planarity
of the proof net.
It is also not at all clear that a version of the matrix method could
somehow be formulated for this kind of logic, for now the specific
arrangement of the formulae is critical.

\subsection{Nonassociative Lambek calculus}

We finally turn to a version of Lambek calculus from which even the
implicit structural rule of associativity is taken away.
This nonassociative Lambek system NL was first described in 1961
\cite{Lambek1961} where it was motivated by linguistic applications, and it has since been recognized as fundamental
within the area of type-logical grammars for linguistics \cite{Moortgat1997,Fulop2004a}. 
This system is especially useful for grammatical deductions because, without
associative sequences, the sequent calculus handles binary trees of formulae
that can be used to represent the syntactic structures of languages.
The sequent presentation below does without the product operator
`$\bullet$', because this is logically superfluous in a sequent
formulation (as it is even in the
associative system L above). 
The nonassociativity of the sequents is here emphasized by replacing
the usual comma with the sequent-level operator `$\diamond$'.
The sequent system enjoys Cut-elimination and is single-conclusion, so
that all provable sequents have a single succedent formula.

\begin{Def}[NL sequent calculus, \cite{Lambek1961}]  \mbox{}\\[-14pt] \label{NL}
\begin{multicols}{2}
\setlength{\abovedisplayskip}{0pt}
\begin{gather*} 
A \Rightarrow A \quad \textup{(Axiom)}\\[10pt]
\genfrac{}{}{}{0}{\Delta \Rightarrow A \quad \Gamma[A] \Rightarrow
C}{\Gamma[\Delta] \Rightarrow C} \quad \textup{(Cut)}\\[10pt]
\genfrac{}{}{}{0}{\Delta \Rightarrow B \quad \Gamma[A] \Rightarrow
C}{\Gamma[(A/B \diamond \Delta)] \Rightarrow C}
\quad \textup{($/$ L)}\\[10pt]
\genfrac{}{}{}{0}{\Delta \Rightarrow B \quad \Gamma[A] \Rightarrow
C}{\Gamma[(\Delta \diamond B\backslash A)] \Rightarrow C} \quad \textup{($\backslash$ L)}\\[10pt]
\genfrac{}{}{}{0}{(\Gamma \diamond B) \Rightarrow A}{\Gamma
\Rightarrow A/B} \quad \textup{($/$ R)}\\[10pt]
\genfrac{}{}{}{0}{(B \diamond \Gamma) \Rightarrow A}{\Gamma
\Rightarrow B\backslash A} \quad \textup{($\backslash$ R)}
\end{gather*} \end{multicols}
\end{Def}

Despite having been discussed many times in the literature, the system
NL has never had a proof net scheme defined for it in a way that
relates clearly with the other proof net schemes presented here.\footnote{A
  proof net system for ``classical'' NL was provided in
  \cite{deGrooteLamarche2002}, but these authors used a quite different
  formulation whose definition and correctness condition bears little
  obvious resemblance to the proof nets so far discussed.}
It turns out to be quite easy to adapt the proof nets for associative
L, but an additional correctness condition is required that has never been developed in literature, and which makes
the resulting nets even more ``geometrical.''

Let us discuss several examples of NL proof structures to develop the additional correctness condition.
Example~1 shows the basic kind of structure for a provable sequent,
which is planar just as in system L.
A further correctness condition is needed in order to account for
the effects of the parentheses, which govern the nonassociative
structure of the antecedent.
To develop this extra condition, we draw dotted boundaries from each pair of parentheses in the
antecedent, extending around the first decomposition link whose active
conclusion subformula is governed by that pair.
Such boundaries in our proof nets will be called \emph{parenthetical boundaries.}
Examining the axiom links in the final structure, observe that only
the link coming from the negative $B$ atom, which connects to the
succedent, crosses the parenthetical boundary that contains it.

\textbf{Example 1}
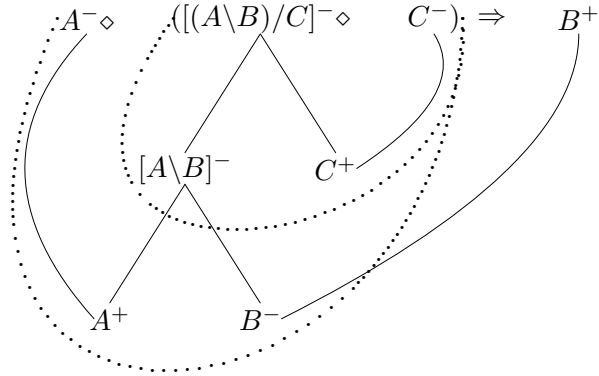
\[
\object{\xybox{ 0*{A^- \diamond}, +/r6em/="AbackB/C"*{([(A\backslash B)/C]^- \diamond}, +/r6em/="C-"*{C^-)}
,+/r2em/*{\Rightarrow}, +/r3em/="B+"*{B^+}
,"AbackB/C"+/d2cm/+/l1cm/="AbackB"*{[A\backslash B]^-}
,"AbackB/C"+/d2cm/+/r1cm/="C+"*{C^+}
,"AbackB"+/d2cm/+/l1cm/="A+"*{A^+}, "AbackB"+/d2cm/+/r1cm/="B-"*{B^-}
,"AbackB"+/u6pt/;"AbackB/C"+/d6pt/**@{-};"C+"+/u6pt/**@{-}
,"A+"+/u6pt/;"AbackB"+/d6pt/**@{-};"B-"+/u6pt/**@{-}
,"A+"+/l6pt/;0+/d6pt/**\crv{"AbackB"+/l3cm/}
,"C+"+/r8pt/;"C-"+/d6pt/**\crv{"C-"+/d1cm/+/r5mm/}
,"B-"+/r8pt/;"B+"+/d6pt/**\crv{"B+"+/d2cm/}
,"AbackB/C"+/l3em/;"C-"+/r1em/**\crv{~*=<4pt>{.} "AbackB"+/d2cm/+/l3cm/&"C+"+/d15mm/+/r2cm/}
,0+/l1em/;"C-"+/r1em/**\crv{~*=<4pt>{.} "A+"+/d25mm/+/l35mm/&"B-"+/d20mm/+/r25mm/}  }}
\]

Example~2, by contrast, shows a similar proof structure for a
non-provable sequent whose antecedent has the parentheses wrong.
The structure is indeed planar, so the sequent would be provable in
system L by invoking associativity, but observe that now both of the antecedent axiom links cross the first
boundary (the outer boundary is not shown).
The problem, in reality, is the $C$-link, because \emph{the link from the
positive atom crosses the parenthetical boundary which contains it.}
We therefore state this as part of the correctness conditions.
\begin{Theorem}
An NL-proof structure is a proof net for the decomposed sequent just in case:
\begin{itemize}
\item the Danos-Regnier switching condition, or other equivalent condition, holds of the structure;
\item the structure is planar;
\item no axiom link from a positive atom crosses the parenthetical boundary which contains it.
\end{itemize}
\end{Theorem}
Example~3 illustrates the structure for a more complicated provable sequent,
and we observe that two axiom links cross boundaries, but neither
involves a link from a positive atom crossing the boundary which
contains it.
 
\textbf{Example 2}
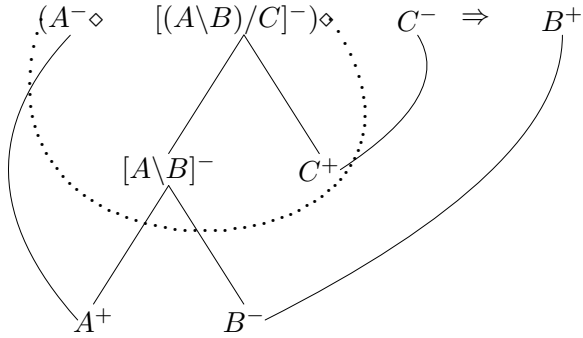
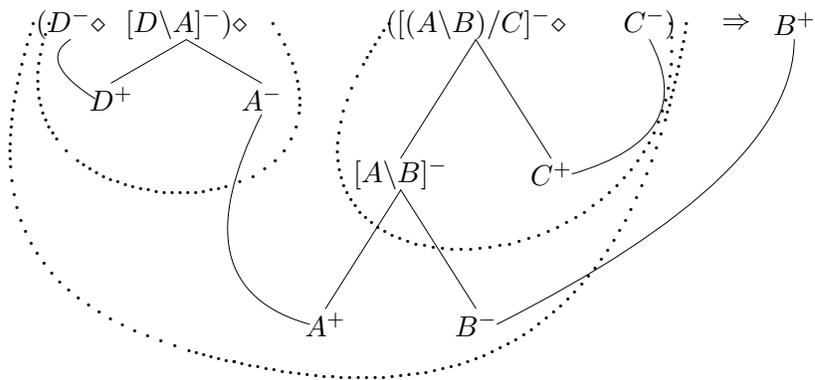
\[
\object{\xybox{ 0*{(A^- \diamond}, +/r6em/="AbackB/C"*{[(A\backslash B)/C]^-) \diamond}, +/r6em/="C-"*{C^-}
,+/r2em/*{\Rightarrow}, +/r3em/="B+"*{B^+}
,"AbackB/C"+/d2cm/+/l1cm/="AbackB"*{[A\backslash B]^-}
,"AbackB/C"+/d2cm/+/r1cm/="C+"*{C^+}
,"AbackB"+/d2cm/+/l1cm/="A+"*{A^+}, "AbackB"+/d2cm/+/r1cm/="B-"*{B^-}
,"AbackB"+/u6pt/;"AbackB/C"+/d6pt/**@{-};"C+"+/u6pt/**@{-}
,"A+"+/u6pt/;"AbackB"+/d6pt/**@{-};"B-"+/u6pt/**@{-}
,"A+"+/l6pt/;0+/d6pt/**\crv{"AbackB"+/l3cm/}
,"C+"+/r8pt/;"C-"+/d6pt/**\crv{"C-"+/d1cm/+/r5mm/}
,"B-"+/r8pt/;"B+"+/d6pt/**\crv{"B+"+/d2cm/}
,0+/l1em/;"AbackB/C"+/r3em/**\crv{~*=<4pt>{.} "AbackB"+/d2cm/+/l3cm/&"C+"+/d15mm/+/r25mm/} }}
\]

\textbf{Example 3}
\[
\object{\xybox{ 0*{(D^- \diamond}, +/r4em/="DbackA"*{[D\backslash A]^-) \diamond}
,+/r10em/="AbackB/C"*{( [(A\backslash B)/C]^- \diamond}
,+/r6em/="C-"*{C^-)}, +/r3em/*{\Rightarrow}, +/r2em/="B+"*{B^+}
,"DbackA"+/d1cm/+/l1cm/="D+"*{D^+}, "DbackA"+/d1cm/+/r1cm/="A-"*{A^-}
,"AbackB/C"+/d2cm/+/l1cm/="AbackB"*{[A\backslash B]^-}
,"AbackB/C"+/d2cm/+/r1cm/="C+"*{C^+}
,"AbackB"+/d2cm/+/l1cm/="A+"*{A^+}, "AbackB"+/d2cm/+/r1cm/="B-"*{B^-}
,"D+"+/u6pt/;"DbackA"+/d6pt/**@{-};"A-"+/u6pt/**@{-}
,"AbackB"+/u6pt/;"AbackB/C"+/d6pt/**@{-};"C+"+/u6pt/**@{-}
,"A+"+/u6pt/;"AbackB"+/d6pt/**@{-};"B-"+/u6pt/**@{-}
,0+/d6pt/;"D+"+/l6pt/**\crv{"D+"+/u5mm/+/l1cm/}
,"C+"+/r8pt/;"C-"+/d6pt/**\crv{"C+"+/u5mm/+/r2cm/}
,"A+"+/l6pt/;"A-"+/d6pt/**\crv{"A+"+/u5mm/+/l2cm/}
,"B-"+/r8pt/;"B+"+/d6pt/**\crv{"B+"+/d2cm/}
,0+/l8pt/;"DbackA"+/r3em/**\crv{~*=<4pt>{.} "D+"+/d2cm/+/l2cm/&"A-"+/d2cm/+/r2cm/}
,"AbackB/C"+/l3em/;"C-"+/r8pt/**\crv{~*=<4pt>{.} "AbackB"+/d2cm/+/l3cm/&"C+"+/d2cm/+/r2cm/}
,0+/l14pt/;"C-"+/r14pt/**\crv{~*=<4pt>{.} "D+"+/l2cm/+/d2cm/&"A+"+/d1cm/+/l2cm/&"B-"+/d1cm/+/r2cm/}
}}
\]

It is quite easy to see the necessity of this correctness condition, so a brief explanation should suffice as proof here.
Note first that for each slash operator in a provable NL sequent, there must be a pair of parentheses surrounding a formula which contains it, and also surrounding the neighboring occurrence of the subformula under the slash.
Every atomic subformula under a top-level slash (i.e.\ one not itself within a proper subformula) in the antecedent of the sequent will decompose to a positive signed atom in the proof net, while the neighboring atom of the same label will show the opposite sign (cf.\ Example~1).
With the parentheses in the right place, an axiom link connecting the two atoms will not cross a boundary determined by them; with parentheses in the wrong places, the positive atom will be contained within a boundary which does not contain its counterpart negative (cf.\ Example~2).
This argument extends by a structural induction to more complicated formulae. 
In essence, the device of the boundaries is a way of checking the grouping action of the parentheses in the proof net.
  
Example~4 shows that care must be taken to draw the parenthetical boundaries when
subformulae involving ``useless types'' occur in the sequent.
Observe that the subformula $C/D$ has types $C$ and $D$ which are
``useless,'' in that they do no work in reducing the sequent.
When this is the case, we must draw the boundary all the way around
the axiom links which connect the decompositions of the occurrences of
$C/D$.
After drawing the boundaries appropriately, we observe that once again
no axiom link coming from a positive atom crosses a boundary which
contains it.
The proof structure for this provable sequent is then a proof net,
under our newly formulated condition.

\textbf{Example 4}
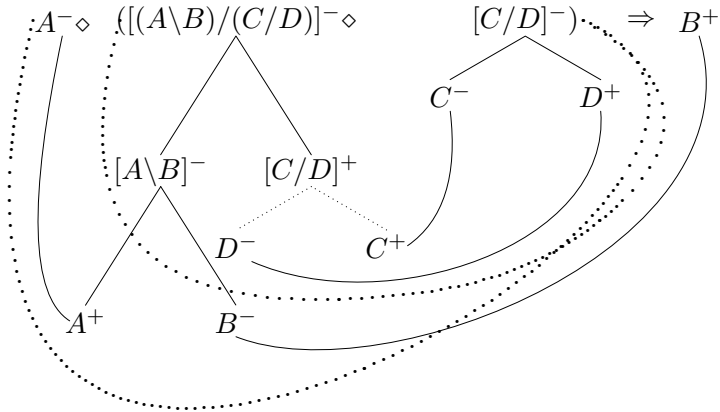
\[
\object{\xybox{ 0*{A^- \diamond},
+/r6em/="crud"*{([(A\backslash B)/(C/D)]^- \diamond},
+/r10em/="C/D-"*{[C/D]^-)}, +/r4em/*{\Rightarrow}, +/r2em/="B+"*{B^+}
,"crud"+/d2cm/+/l1cm/="AbackB"*{[A\backslash B]^-}
,"crud"+/d2cm/+/r1cm/="C/D+"*{[C/D]^+}
,"C/D-"+/d1cm/+/l1cm/="C-"*{C^-}, "C/D-"+/d1cm/+/r1cm/="D+"*{D^+}
,"AbackB"+/d2cm/+/l1cm/="A+"*{A^+}, "AbackB"+/d2cm/+/r1cm/="B-"*{B^-}
,"C/D+"+/d1cm/+/l1cm/="D-"*{D^-}, "C/D+"+/d1cm/+/r1cm/="C+"*{C^+}
,"AbackB"+/u6pt/;"crud"+/d6pt/**@{-};"C/D+"+/u6pt/**@{-}
,"D-"+/u6pt/;"C/D+"+/d6pt/**@{.};"C+"+/u6pt/**@{.}
,"A+"+/u6pt/;"AbackB"+/d6pt/**@{-};"B-"+/u6pt/**@{-}
,"C-"+/u6pt/;"C/D-"+/d6pt/**@{-};"D+"+/u6pt/**@{-}
,0+/d6pt/;"A+"+/l6pt/**\crv{"A+"+/u5mm/+/l1cm/}
,"C+"+/r8pt/;"C-"+/d6pt/**\crv{"C+"+/u5mm/+/r1cm/}
,"D-"+/r6pt/+/d6pt/;"D+"+/d6pt/**\crv{"C+"+/d1cm/&"C+"+/r3cm/}
,"B-"+/d6pt/;"B+"+/d6pt/**\crv{"C+"+/d2cm/&"C+"+/r5cm/}
,"crud"+/l4em/;"C/D-"+/r2em/**\crv{~*=<4pt>{.} "AbackB"+/l15mm/+/d15mm/&"C+"+/d1cm/&"C+"+/r3cm/&"D+"+/r15mm/}
,0+/l1em/;"C/D-"+/r2em/**\crv{~*=<4pt>{.} "A+"+/l20mm/+/d15mm/&"C+"+/d3cm/&"D+"+/r20mm/}
}}
\]

We have at last descended all the way down the substructural
hierarchy.
As the logics became more stringent in dealing with a specific
arrangement of formulae, the conditions on proof nets became
accordingly more geometrical.
Moreover, we noted that for those logics that do not deal with a
specific arrangement of the formulae, it was possible to invoke the
extremely compressed proof format of the matrix method.
It is the author's hope that this unified discussion has illuminated
the ways in which the ``geometry'' required of formulae in a logical
consequence relation ends up being encoded into the ``geometry of
proofs'' validating sequents for the logic.
There are probably also some connections that could now be made with work that has explicitly represented proof nets topologically as cell complexes (e.g.\ \cite{Metayer2001}).

\newpage

\bibliographystyle{asl}
\bibliography{/home/safulop/bibfiles/math,/home/safulop/bibfiles/logic,/home/safulop/bibfiles/alglogic,/home/safulop/bibfiles/mathling,/home/safulop/bibfiles/Fulop}

\end{document}